# Grand potential in thermodynamics of solid bodies and surfaces

**A.I. Rusanov**
*Mendeleev Center, St. Petersburg State University, St. Petersburg 199034, Russia*

**A.K. Shchekin, and D.V. Tatyanenko**
*Department of Statistical Physics, Faculty of Physics, St. Petersburg State University, Ulyanovskaya 1, Petrodvorets, St. Petersburg 198504, Russia*

Using the chemical potential of a solid in a dissolved state or the corresponding component of the chemical potential tensor at equilibrium with the solution, a new concept of grand thermodynamic potential for solids has been suggested. This allows generalizing the definition of Gibbs' quantity $\sigma$ (surface work often called the solid-fluid interfacial free energy) at a planar surface as an excess grand thermodynamic potential per unit surface area that (1) does not depend on the dividing surface location and (2) is common for fluids and solids.

In its development (see, e.g., surveys[1,2]), the surface thermodynamics of solids seems to become more and more complicated as compared with the thermodynamics of fluids. Concerning solid surfaces, Gibbs (p. 315 in Ref. 3) introduced his famous quantity $\sigma$ as the work of formation of a new surface per unit area and was first to distinguish between $\sigma$ and the mechanical surface tension as an excess surface stress. Gibbs used $\sigma$ in all his formulas as referring to the equimolecular dividing surface that practically coincides with the boundary surface of a solid. Gibbs himself did not consider the dependence of $\sigma$ on the dividing surface location. Only recently this question was touched upon by Schimmele et al.[4] in a paper devoted to line tension. As for mechanical surface tension, it generally is a tensorial quantity, $\hat{\gamma}$, and, only in the isotropic case (when $\hat{\gamma} = \gamma\hat{1}$, $\hat{1}$ being the unit tensor), is reduced to a scalar quantity $\gamma$ that can be compared with $\sigma$. As is well known, $\gamma = \sigma$ for planar fluid surfaces in the absence of an external field,[5,6] but $\gamma$ and $\sigma$ are different from each other for solids. As was first explained by one of us,[7] the cause of a difference between $\sigma$ and the mechanical surface tension lies in the presence of a non-uniformity of the chemical potential of matter in the surface layer of a solid. To be more exact, one has to speak about the immobile component of a solid that forms a solid lattice. An immobile component is incapable of diffusion, so that the main mechanism of leveling chemical potentials does not work for immobile components in solids. Gibbs' quantity $\sigma$ is often called surface free energy in the literature, but this definition is strict only for a specific dividing surface coinciding with the equimolecular surface in a single-component system[1] (pure substance). Below, we generalize the definition on the basis of grand thermodynamic potential.

For a fluid system, grand thermodynamic potential $\Omega$ is defined as

$$\Omega \equiv F - \sum_i \mu_i N_i, \tag{1}$$

where $F$ is the free energy of the system and $\mu_i$ and $N_i$ are the chemical potential and the number of molecules of the $i$th species, respectively. Correspondingly, $\sigma$ (coinciding this time with surface tension) is defined as

$$\sigma = \bar{\Omega}/A, \tag{2}$$

where the bar denotes a surface excess value and $A$ is the dividing surface area. For fluids, surface tension is known to be independent of the dividing surface location in the case of a planar interface, so that, naturally, Eq. (2) is valid at an arbitrary choice of the dividing surface. Since the chemical potential of an immobile component of a solid (which is a tensor) ceases to be a variable in the grand canonical ensemble, grand thermodynamic potential seems to be inapplicable to solids. Nevertheless, it is a common practice to use grand thermodynamic potential when considering open complex systems including both fluid and solid phases (e.g., wetting systems). To validate this practice, a hybrid potential $\tilde{\Omega}$ was introduced[7] as defined again according to Eq. (1) but with subscript $i$ referring to mobile components only. Such definition implies that $\tilde{\Omega}$ behaves as free energy with respect to immobile components (in solids) and as grand thermodynamic potential with respect to mobile components (both in solids and fluids).

Assuming the presence of only a single immobile component $j$ (the restriction being of no principal significance), fundamental equations in terms of $\tilde{\Omega}$ for a uniform solid phase are[1,2]

$$d\tilde{\Omega} = -SdT + \hat{E} : d\hat{V} + \hat{\mu}_j : d\hat{N}_j - \sum_i N_i d\mu_i , \qquad (3)$$

$$\tilde{\Omega} = E_{xx}V + \mu_{j(xx)}N_j , \qquad (4)$$

where $S$ is entropy; $T$ is temperature; $\hat{E}$ is the stress tensor; $\hat{V}$ is the volume displacement tensor;[2] colon denotes a biscalar product (the convolution) of tensors; $\hat{\mu}_j$ and $\hat{N}_j$ are the chemical potential and the mass displacement tensors,[2] respectively, for the immobile component; $E_{xx}$ and $\mu_{j(xx)}$ are respective diagonal components of $\hat{E}$ and $\hat{\mu}_j$ for an arbitrary direction $x$ in which the solid phase is uniform; $V$ is the phase volume, and $N_j$ is the number of molecules forming the crystalline lattice. The volume displacement tensor trivially differs from the strain tensor by showing not linear but volume changes at strain, and, similarly, the mass displacement tensor shows mass changes when the crystal grows (or dissolves/evaporates) in various directions.[2] Equation (4) is a result of integration of Eq. (3) along a selected direction $x$ at fixed physical state. Using the hybrid potential, $\sigma$ is defined as an excess quantity of $\tilde{\Omega}$ per unit surface area for the particular location of a dividing surface where the adsorption of the immobile component is zero,[1] which exactly corresponds to Gibbs' approach.

The goal of this communication is formulating an alternative approach on the basis of grand thermodynamic potential specified for solids with participation of all species. This will lead to more general consideration including a shift of a dividing surface and a universal definition of Gibbs' quantity $\sigma$ valid for an arbitrary location of the dividing surface.

Every solid contains at least one immobile component by necessity. In addition, a solid can also contain mobile components that freely move inside the solid lattice. Gibbs called them "fluid components" (fluids absorbed by solids) and showed them to behave as in pure fluid systems including the definition and uniformity of chemical potentials at equilibrium (p. 216 in Ref. 3). Although the presence of mobile components in a solid is quite unnecessary (and is not typical for practice), we, for the sake of generality and following Gibbs, will address the case when both the types of components occur in a solid. Analyzing the solubility, we are naturally interested, first of all, in studying the behavior of an immobile component that can be soluble or insoluble in an adjacent fluid.

We first consider the case when a solid (phase $\alpha$) is in a real equilibrium with a fluid phase $\beta$ where the solid is partly soluble. The solid is assumed to be macroscopic, which allows us to consider the solid/fluid interface as flat. Let us introduce Cartesian coordinates $x$, $y$, $z$ with the $z$-axis normal to the interface and directed from phase $\alpha$ to phase $\beta$. For the sake of simplicity, we assume no shear stresses in the solid, so that the coordinate axes simultaneously are the principal directions for the stress tensor in the solid. This is just the case that was analyzed by Gibbs (page 194 in Ref. 3) who derived the equilibrium condition (Gibbs' Eq. (661))

$$(f^\alpha - E_z^\alpha - \sum_i \mu_i^\alpha c_i^\alpha)/c_j^\alpha = \mu_j^\beta, \tag{5}$$

where $f$ is the free energy density, $E_z$ is the principal value of the stress tensor $\hat{E}$ corresponding to direction $z$ as a normal to the interface, $c$ is the concentration (the number of molecules per unit volume), subscripts $i$ and $j$ refer to mobile and immobile components, respectively. Rigorously defining the chemical potential tensor $\hat{\mu}_j^\alpha$ for a solid,[1,2] the left-hand side of Eq. (5) can be identified with the principal value $\mu_{j(z)}^\alpha$ of tensor $\hat{\mu}_j^\alpha$ to yield the phase equilibrium condition

$$\mu_{j(z)}^\alpha = \mu_j^\beta. \tag{6}$$

Naturally, the mechanical equilibrium condition should be fulfilled in addition

$$E_z^\alpha = -p^\beta, \tag{7}$$

where $p^\beta$ is the hydrostatic pressure in fluid phase $\beta$.

In the situation we consider, soluble species $j$ simultaneously plays the role of an immobile component (in phase $\alpha$) and a mobile component (in phase $\beta$). Therefore, we now may regard the chemical potential $\mu_j^\beta$ as a variable of the grand canonical ensemble (e.g. by allowing the system to be in contact with a large reservoir of phase $\beta$). Let us define the grand thermodynamic potential of the two-phase system under consideration as

$$\Omega \equiv F - \sum_i \mu_i N_i - \mu_j^\beta N_j = \tilde{\Omega} - \mu_j^\beta N_j. \tag{8}$$

Applying this definition to the homogeneous bulk phases and taking into account Eq. (4) yield

$$\Omega^\alpha = E_k^\alpha V^\alpha + (\mu_{j(k)}^\alpha - \mu_j^\beta)N_j^\alpha, \quad k = x, y, z, \tag{9}$$

$$\Omega^\beta = -p^\beta V^\beta, \tag{10}$$

or, passing to the density of grand thermodynamic potential $\omega \equiv \Omega/V$,

$$\omega^\alpha = E_k^\alpha + (\mu_{j(k)}^\alpha - \mu_j^\beta)c_j^\alpha, \quad k = x, y, z, \tag{11}$$

$$\omega^\beta = -p^\beta. \tag{12}$$

For the density of hybrid potential $\tilde{\omega} \equiv \tilde{\Omega}/V$, we have the following expressions

$$\tilde{\omega}^\alpha = E_k^\alpha + \mu_{j(k)}^\alpha c_j^\alpha, \quad k = x, y, z, \tag{13}$$

$$\tilde{\omega}^\beta = -p^\beta + \mu_j^\beta c_j^\beta. \tag{14}$$

Since energetic quantities $\omega^\alpha$ and $\tilde{\omega}^\alpha$ cannot depend on a direction in the bulk phase, the right-hand sides of Eqs. (11) and (13) should be invariant with respect to a principal direction. This implies the equalities

$$E_x^\alpha + \mu_{j(x)}^\alpha c_j^\alpha = E_y^\alpha + \mu_{j(y)}^\alpha c_j^\alpha = E_z^\alpha + \mu_{j(z)}^\alpha c_j^\alpha. \tag{15}$$

The above expressions refer to a bulk phase. Passing now to an interface and introducing the interfacial local concentration $c_j(z)$, the interfacial local densities of grand thermodynamic potential $\omega(z)$ and hybrid potential $\tilde{\omega}(z)$, we write, by analogy with Eqs. (11) and (13),

$$\omega(z) = E_k(z) + [\mu_{j(k)}(z) - \mu_j^\beta] c_j(z), \quad k = x, y, \tag{16}$$

$$\tilde{\omega}(z) = E_k(z) + \mu_{j(k)}(z) c_j(z), \quad k = x, y. \tag{17}$$

The case $k = z$ is automatically excluded in Eqs. (16) and (17) since thermodynamic integral expressions can be written only for directions in which a system is uniform (e.g., when passing from Eq. (3) to Eq. (4)). In our case of a two-phase system, the system is uniform in directions $x$ and $y$ at any fixed value of $z$, but is inhomogeneous in direction $z$ at the interface. To derive excess quantities $\bar{\omega} \equiv \bar{\Omega}/A$ for grand thermodynamic potential and $\bar{\tilde{\omega}} \equiv \bar{\tilde{\Omega}}/A$ for hybrid potential, we introduce a dividing surface with coordinate $z_0$ and write

$$\bar{\omega} = \int_{-\infty}^{z_0} [\omega(z) - \omega^\alpha] dz + \int_{z_0}^{\infty} [\omega(z) - \omega^\beta] dz, \tag{18}$$

$$\bar{\tilde{\omega}} = \int_{-\infty}^{z_0} [\tilde{\omega}(z) - \tilde{\omega}^\alpha] dz + \int_{z_0}^{\infty} [\tilde{\omega}(z) - \tilde{\omega}^\beta] dz, \tag{19}$$

or, putting here Eqs. (11)–(17),

$$\bar{\omega} = \gamma_k + \int_{-\infty}^{z_0} [(\mu_{j(k)}(z) - \mu_j^\beta) c_j(z) - (\mu_{j(k)}^\alpha - \mu_j^\beta) c_j^\alpha] dz + \int_{z_0}^{\infty} (\mu_{j(k)}(z) - \mu_j^\beta) c_j(z) dz, \quad k = x, y, \tag{20}$$

$$\bar{\tilde{\omega}} = \gamma_k + \int_{-\infty}^{z_0} [\mu_{j(k)}(z) c_j(z) - \mu_{j(k)}^\alpha c_j^\alpha] dz + \int_{z_0}^{\infty} [\mu_{j(k)}(z) c_j(z) - \mu_j^\beta c_j^\beta] dz, \quad k = x, y, \tag{21}$$

where

$$\gamma_k = \int_{-\infty}^{z_0} (E_k - E_k^\alpha) dz + \int_{z_0}^{\infty} (E_k + p^\beta) dz, \quad k = x, y \tag{22}$$

are the principal values of the two-dimensional tensor of mechanical surface tension. Adding and subtracting $\mu_{j(k)}^\alpha c_j(z)$ in the integrand of the first integrals in Eqs. (20) and (21), adding and subtracting $\mu_j^\beta c_j(z)$ in the integrand of the second integral in Eq. (21), we rearrange Eqs. (20) and (21) to the form

$$\bar{\omega} = \gamma_k + \bar{g}_{j(k)} + (\mu_{j(k)}^\alpha - \mu_j^\beta) \Gamma_j^\alpha, \quad k = x, y, \tag{23}$$

$$\bar{\tilde{\omega}} = \gamma_k + \bar{g}_{j(k)} + \mu_{j(k)}^\alpha \Gamma_j^\alpha + \mu_j^\beta \Gamma_j^\beta, \quad k = x, y, \tag{24}$$

where

$$\bar{g}_{j(k)} = \int_{-\infty}^{z_0} (\mu_{j(k)}(z) - \mu_{j(k)}^\alpha) c_j(z) dz + \int_{z_0}^{\infty} (\mu_{j(k)}(z) - \mu_j^\beta) c_j(z) dz, \quad k = x, y, \tag{25}$$

and the quantities

$$\Gamma_j^\alpha \equiv \int_{-\infty}^{z_0} (c_j(z) - c_j^\alpha) dz, \quad \Gamma_j^\beta \equiv \int_{z_0}^{\infty} (c_j(z) - c_j^\beta) dz \tag{26}$$

are the adsorptions of the immobile component on the sides of phase $\alpha$ and phase $\beta$, respectively.

Let us now show that excess quantity $\bar{\omega}$ does not depend at all on the dividing surface location. To do this, we differentiate Eq. (18) with respect to $z_0$ at a fixed physical state. This yields

$$\left[d\bar{\omega}/dz_0\right] = -\omega^\alpha + \omega^\beta, \qquad (27)$$

where brackets indicate the derivative to be of no physical sense and only to correspond to an imaginary shift of a dividing surface. Putting Eqs. (11) and (12) in Eq. (27) results in

$$\left[d\bar{\omega}/dz_0\right] = -E_k^\alpha - (\mu_{j(k)}^\alpha - \mu_j^\beta)c_j^\alpha - p^\beta, \quad k = x, y. \qquad (28)$$

Since the right-hand side of Eq. (28) contains only quantities referring to the bulk phases, we may use the condition expressed in Eq. (15) to obtain

$$\left[d\bar{\omega}/dz_0\right] = -E_z^\alpha - (\mu_{j(z)}^\alpha - \mu_j^\beta)c_j^\alpha - p^\beta. \qquad (29)$$

Applying now the equilibrium conditions expressed in Eqs. (6) and (7), we arrive at the conditions

$$\left[d\bar{\omega}/dz_0\right] = -\omega^\alpha + \omega^\beta = 0. \qquad (30)$$

Thus, we obtained the same equilibrium conditions as for two fluid phases in equilibrium, but with our newly introduced grand thermodynamic potential of solids. The condition $\omega^\alpha = \omega^\beta$ can be of significance in the theory of wetting.[4]

Let us now consider the case of an arbitrary location of the dividing surface. Gibbs himself defined the quantity $\sigma$ only for the dividing surface equimolecular with respect to the immobile component of solid. However, he also considered a physical interpretation of $\sigma$ as the work of formation of unit new surface, e.g. in cutting up the body (see p. 315 in Ref. 3). Let us accept this as the basic definition to be applied to an arbitrary dividing surface. We will follow the scheme used in Ref. 1 (see p. 192 therein) but employing the grand potential instead of the hybrid one.

Considering the process of formation of a new surface of a solid at a given temperature, total volume of the system and chemical potentials $\{\mu_i^\beta\}$ and $\mu_j^\beta$ in the adjacent fluid phase, we may calculate the work of the process as the difference between the final and the initial values of the grand potential $\Omega$ defined by Eq. (8). Let a rectangular-prism-shaped piece of the solid (phase α) of volume $V^\alpha$ be in the fluid medium (phase β) in thermodynamic equilibrium with it; the whole system is considered in a certain fixed volume. Cutting up the solid into two pieces (both remaining in our fixed volume) requires a work

$$\Delta\Omega = (\omega^\alpha - \omega^\beta)\Delta V^\alpha + \bar{\omega}\Delta A \qquad (31)$$

with $\Delta V^\alpha$ the change of the volume of the solid (possible due to choice of the dividing surfaces) and $\Delta A$ the change of the area of the solid surface. Implying the solid to be cut by the plane parallel to one of its faces, we see that $\Delta A$ does not depend on choice of the dividing surfaces at the cut. Using the condition (30), we then obtain $\Delta\Omega = \bar{\omega}\Delta A$. By Gibbs' definition of $\sigma$ this work equals $\sigma\Delta A$. Therefore, $\sigma = \bar{\omega}$ for an arbitrary solid–fluid dividing surface.

Comparing this equality with Eq. (23) we obtain the relation (cf. (Eq. (5.31) in Ref. 1)

$$\sigma = \gamma_k + \bar{g}_{j(k)} + (\mu_{j(k)}^\alpha - \mu_j^\beta)\Gamma_j^\alpha, \quad k = x, y. \qquad (32)$$

As follows from scalar sense of the quantity $\sigma$, the right-hand side of Eq. (32) is an invariant with respect to the choice of direction along the surface. As one can see from Eq. (32), the difference between the mechanical surface tension and Gibbs' quantity $\sigma$ for solids is caused by the two terms. The term $\bar{g}_{j(k)}$ characterizes non-uniformity of the chemical potential of the immobile component. The other term, $(\mu_{j(k)}^\alpha - \mu_j^\beta)\Gamma_j^\alpha$, is non-zero at both non-zero adsorption $\Gamma_j^\alpha$ and an anisotropy of the chemical potential of the solid bulk. Indeed, applying the condition (6) of chemical equilibrium to an isotropic state of the solid bulk, we obtain $\mu_{j(x)}^{\alpha} = \mu_{j(y)}^{\alpha} = \mu_{j(z)}^{\alpha} = \mu_{j}^{\beta}$. Thus, in this case the term $(\mu_{j(k)}^{\alpha} - \mu_{j}^{\beta})\Gamma_{j}^{\alpha}$ is zero at arbitrary choice of the dividing surface.

We now can say that if we return to the definition (2) of $\sigma$ as specific excess grand thermodynamic potential $\bar{\omega}$ and apply this definition not only to fluids, but also to solids, such definition will be valid for an arbitrary position of the dividing surface provided the grand thermodynamic potential is defined according to Eq. (8). As for the hybrid potential, its excess surface density, Eq. (24), does depend on the dividing surface location and can be related to $\sigma$ only at a special choice of a dividing surface.[1] Since Eqs. (23) and (24) imply $\bar{\omega} = \bar{\tilde{\omega}} - \mu_{j}^{\beta}(\Gamma_{j}^{\alpha} + \Gamma_{j}^{\beta})$ and, as we have shown, always $\bar{\omega} = \sigma$, we obtain that, for a non-zero $\mu_{j}^{\beta}$, this dividing surface is necessarily equimolecular with respect to the immobile component $j$ ($\Gamma_{j}^{\alpha} + \Gamma_{j}^{\beta} = 0$). So, generalizing the definition of Gibbs' quantity $\sigma$ with the aid of the newly introduced grand thermodynamic potential of solids, we obtain a more convenient scientific tool, not to speak of the commonality of such definition for fluids and solids.

The discussion above referred to the case when a soluble solid is in equilibrium with a real fluid. For the case when a solid is insoluble in a given fluid or an adjacent fluid is absent at all, a more general approach can be elaborated using a hypothetical equilibrium of the solid in an arbitrary state with an imaginary fluid where the chemical potential of the component $j$ is just equal to $\mu_{j(k)}^{\alpha}$. Implying the equilibrium condition expressed in Eq. (6), we can define the grand thermodynamic potential of a solid as

$$\Omega \equiv F - \sum_{i}\mu_{i}N_{i} - \mu_{j(z)}^{\alpha}N_{j}, \qquad (33)$$

which is replacing $\mu_{j}^{\beta}$ by $\mu_{j(k)}^{\alpha}$ in Eq. (8). Equation (33) has no reference to a co-existing fluid phase, and the whole theory is formulated in a general form. With the replacement of $\mu_{j}^{\beta}$ by $\mu_{j(k)}^{\alpha}$, all the above relationships are maintained, as well as a universal definition of Gibbs' quantity $\sigma$.

Two concluding remarks can be given. First, our theory was formulated for uniform bulk phases. Passing to non-uniform phases, especially often met for solids in practice, all the above equations should be considered as local relationships. Second, we confined our analysis with only flat surfaces. The case of a curved solid surface is complicated not only by a stress jump, but also by non-equality of chemical potentials in adjacent phases. This case remains to be a challenge for the theory of surfaces, although first steps in its analysis has already been done.[8]

This work was supported by the Program "The Development of Scientific Potential of Higher Education" (project RNP.2.1.1.4430). D.V. Tatyanenko thanks the Russian Foundation for Basic Research (grant No. 09-03-01005-a).

[1] A.I. Rusanov, Thermodynamics of solid surfaces, *Surf. Sci. Rep.* **23** (1996) 173.
[2] A.I. Rusanov, Surface thermodynamics revisited, *Surf. Sci. Rep.* **58** (2005) 111.
[3] J.W. Gibbs, The Scientific Papers, Longmans, New York, 1906, 1928.
[4] L. Schimmele, M. Napiórkowski, and S. Dietrich, *J. Chem. Phys.* **127** (2007) 164715; e-print arXiv: cond-mat/0703821.
[5] A.I. Rusanov, F.M. Kuni, *J. Colloid Interface Sci.* **100** (1984) 264.
[6] F.M. Kuni, A.K. Shchekin, A.I. Rusanov, *Colloid Journal of the USSR* **45** (1983) 598.
[7] A.I. Rusanov, *J. Colloid Interface Sci.* **63** (1978) 330.
[8] A.I. Rusanov, A.K. Shchekin, *J. Chem. Phys.* **127** (2007) 191102.